\documentstyle[aaspp4]{article}

\def\msun{M_\odot}

\def\lsun{L_\odot}
\def\fun#1#2{\lower3.6pt\vbox{\baselineskip0pt\lineskip.9pt
  \ialign{$\mathsurround=0pt#1\hfil##\hfil$\crcr#2\crcr\sim\crcr}}}
\def\la{\mathrel{\mathpalette\fun <}}
\def\ga{\mathrel{\mathpalette\fun >}}
\def\eg{{\it e.g., }}
\def\etal{{\it et al. }}

\def\mpc{{\rm Mpc}}

\def\Mpc{{\rm Mpc}}

\def\he#1{\hbox{${}^{#1}{\rm He}$}}
\def\li#1{\hbox{${}^{#1}{\rm Li}$}}

\def\imfm{\xi_{\star}}
\def\mrem{m_{\rm rem}}
\def\avg#1{\langle #1 \rangle}
\def\sigbar{\avg{\sigma}}
\def\rbar{\hbox{$\avg{r}$}}
\def\omegam{\Omega_{\rm Macho}}
\def\omegab{\Omega_{\rm B}}
\def\omegalya{\Omega_{{\rm Ly}\alpha}}

\def\lya{Ly$\alpha$}

\def\pcite#1{(\cite{#1})}
\def\pref#1{(\ref{#1})}

\def\bline{\rule[1.2mm]{3em}{0.1mm}}

\begin{document}

%\slugcomment{{\it New Astronomy}, in press}
%\slugcomment{UMN-TH-1620/98}
\rightline{{\it New Astronomy}, in press}
\rightline{UMN-TH-1620/98}

\title{Massive Compact Halo Objects Viewed from
a Cosmological Perspective:
Contribution to the Baryonic Mass Density of the Universe}

\author{Brian D. Fields}
\affil{University of Minnesota, School of Physics and Astronomy \\
Minneapolis, Minnesota 55455, USA \\
{\tt fields@mnhepw.hep.umn.edu}}

\author{Katherine Freese}
\affil{University of Michigan, Department of Physics \\
Ann Arbor, Michigan 48109-1120, USA \\
{\tt freese@mich.physics.lsa.umich.edu}}

\author{David S. Graff}
\affil{DAPNIA/SPP, CEA Saclay \\
91191 Gif-sur-Yvette cedex, France \\
{\tt graff@hep.saclay.cea.fr}}

\begin{abstract}
We estimate the contribution of Massive Compact Halo Objects (Machos) 
and their stellar progenitors
to the mass density of the Universe.
If the Machos that have been detected reside
in the Halo of our Galaxy, then
a simple extrapolation of the Galactic
population (out to 50 kpc)
of Machos to cosmic scales gives a cosmic density 
$\rho_{\rm Macho} = (1-5) \times 10^9 \, h \, \msun \, \Mpc^{-3}$,
which in terms of the critical density corresponds to 
$\omegam=(0.0036-0.017) h^{-1}$.  
Should the Macho Halo extend
out to much further than 50 kpc, then $\omegam$ would only be larger.
Such a mass density
is comparable to the cosmic baryon density implied by
Big Bang Nucleosynthesis.  If we take
the central values of the estimates, then
Machos dominate the baryonic content of the Universe
today, with  $\omegam/\omegab \sim 0.7 \, h$.  
However, the cumulative uncertainties in 
the density determinations only require that 
$\omegam/\omegab  \geq {1 \over 6} h f_{\rm gal}$,
where the fraction of galaxies that contain Machos
$f_{\rm gal} > 0.17$ and
$h$ is the Hubble constant in units of 100 km s$^{-1}$ Mpc$^{-1}$.
Our best estimate for $\omegam$ is hard to reconcile with the 
current best estimates of the baryonic
content of the intergalactic medium indicated by measurements
of the Lyman-$\alpha$ forest;
however, measurements of $\Omega_{{\rm Ly} \alpha}$
are at present uncertain, so that such a comparison
may be premature.

If the Machos are white dwarfs resulting from a single
burst of star formation (without recycling), then their
main sequence progenitors would have been 
at least twice more massive:
$\Omega_\star=(0.007 - 0.034) h^{-1}$.  
Thus, far too much gaseous baryonic material would remain in the
Galaxy unless there is a Galactic wind to eject it.
Indeed a Macho population of white dwarfs
and the gas ejected from their main sequence progenitors
accounts for a significant fraction of all baryons.  This fact must be taken
into account when attempting to dilute the chemical by-products of
such a large population of intermediate mass stars.
We stress the difficulty of reconciling the Macho mass budget with
the accompanying carbon production in the case of white dwarfs.
In the simplest picture,
even if the excess carbon is ejected from the Galaxy by a Galactic
wind, measurements of carbon abundances in Lyman $\alpha$
forest lines with values $10^{-2}$ solar require that only about
$10^{-2}$ of all baryons can have passed through the white
dwarf progenitors. Such a fraction can barely be accommodated
by our estimates of $\omegam$ and would be in conflict with $\Omega_\star$.

\end{abstract}


\newpage

\section{Introduction}
\label{intro}

In recent years, microlensing experiments (Alcock et al. \cite{macho:2yr};
Renault \cite{ren})
have reported evidence
for Massive Compact Halo Objects (Machos) in the halo of our
Galaxy. Candidates that have been considered as explanations
for these events include faint stars, brown dwarfs, white dwarfs,
and black holes. Other than the possibility of primordial
black holes, all of these candidates are made of baryons.
If these results are
correct, then there may be a significant baryonic component
of the Galactic halo.  Although a 100\% baryonic halo does not
seem likely at this point, as much as 30-50\% of the halo may
be made of baryonic dark matter in the form of Machos 
(Alcock et al. \cite{macho:2yr}).

It is the purpose of this paper to discuss the mass budget
associated with such a discovery, by situating
the discovery of Machos in a cosmological context.
We will obtain a lower limit on
the contribution of Machos to the mass density
of the universe, and compare that contribution with
the baryonic component in the Lyman-$\alpha$ forest and
with the total baryonic mass density allowed by nucleosynthesis.
We will see that, regardless of what the Machos are,
$\omegam$ is comparable
to $\omegab$ (where B refers to baryons).
As an extreme lower limit, we find that 
$\omegam > {h \over 6}  \omegab f_{\rm gal}$,
where $f_{\rm gal}$ is the fraction of galaxies that contain Machos
and $h$ is the Hubble constant in units of 100 km s$^{-1}$ Mpc$^{-1}$.
Taking the central values
for $\omegam$ and $\omegab$, we find that Machos are the
dominant component of baryons in the universe.
It is a remarkable consequence of a mere 6 events that they
could imply such a large fraction of the baryonic mass
budget of the universe.
(At the Aspen Workshop on Microlensing in June 1997,
Tim Axelrod presented preliminary results from the Macho group's 
four year data, 
where there are about 14 events--see also Cook et al.\ \cite{cooketal};
these new numbers do not significantly
change the conclusions of this paper.)

Then we turn to various candidates for the Machos, and
examine the consequences for the mass budget of the universe.
First we consider stellar and substellar candidates such
as brown dwarfs, red dwarfs and ordinary stars.
Next we examine stellar remnants: white dwarfs, neutron stars,
and black holes.  We focus especially on white dwarfs,
as these have been identified as the most popular candidate
because of the best-fit Macho mass of 0.5 $M_\odot$.
With the hypothesis 
that the Machos are white dwarfs, we find a na{\"{\i}}ve
estimate of the contribution of the progenitors to the mass density
of the universe, i.e., we find
$\Omega_\star$ (where $\star$ refers to progenitors).  
Accounting for the progenitors increases the initial
Galactic gas mass needed to give the present Macho observations.
In fact, far too much baryonic material would remain in
the Galaxy unless there is a Galactic wind to eject it.
If the white dwarfs arise from a single burst of star formation,
we find that the main sequence progenitor mass density was
at least twice as large as the Macho mass density.  We examine
the range of overlap between $\Omega_\star$ and $\omegab$.
We also compare with the baryonic content of the intergalactic
medium indicated by measurements of the Lyman-$\alpha$ forest.
{}From these baryon abundance arguments alone,
we will see that white dwarfs survive
as viable Macho candidates, but only if one takes extreme
values of the parameters.
In addition, as we show below, the mass budget discussed in this paper
is difficult to reconcile with the carbon abundance
that would be produced by white dwarf progenitors
(see also Gibson \& Mould \cite{gm}).  

\section{Contribution of Machos to the Mass Density of the
Universe}
\label{naive}

In this section, we make an extrapolation of the microlensing results 
to a cosmological context.  As we argue below, we feel
that this extrapolation provides a robust lower limit
on the Macho contribution to the mass density of the universe.
Microlensing results (Alcock \etal \cite{macho:2yr}) 
predict that the total mass of 
Machos in the Galactic halo out to 50 kpc is 
\begin{equation}
\label{macho_mass}
M_{\rm Macho} = (1.3-3.2) \times 10^{11} \ \msun \, .
\end{equation}
This result does not strongly depend on the halo 
model adopted.  
However, even with a fairly secure value of $M_{\rm Macho}$,
one can only parlay it into a 
cosmological constraint by making
some assumption about the cosmic distribution of Machos.
This in turn requires one to commit at some level
to a scheme 
for the origin and nature of Machos; we will
make the different assumptions explicit as we proceed.
Should the Halo extend to beyond the LMC at 50 kpc,
the numbers in Eq.\ (\ref{macho_mass}) may provide a lower limit
to the total Macho mass in the Halo.  Hence our results
give a lower limit to $\Omega_{Macho}$; still the values
are quite large.

In order to determine a lower limit for the contribution
of Machos to the mass density of the universe, we will
take the following approach.  We can obtain a
``Macho-to-light ratio" for the Machos in the
Galactic Halo; then assuming
that the Macho content of our Galaxy is typical of the
rest of the luminous universe\footnote{We comment on this 
assumption below.  For example, one might posit 
that Macho production is a precursor or byproduct of star formation,
so that Machos exist in the neighborhood of luminous matter and
the result in Eq.\ \pref{omega} is quite accurate.  Alternatively
one can interpret the result as  a lower limit.  Below we discuss how
representative our Galaxy is of the objects contributing to the
luminosity function.},
we can multiply by the average luminosity
density of the universe to determine $\omegam$.  
Thus, if one can determine the ``Macho-to-light'' ratio
for the Milky Way, then one can deduce their cosmic
density.

The total luminosity of the Milky Way is nontrivial
to determine because we are embedded in the Galaxy.
One must rely on Galactic luminosity models
extrapolated beyond the solar neighborhood,
and/or comparisons to similar external galaxies.
Estimates based on Galactic values give 
B-band results $L_{{\rm MW},B} \simeq 1.4\times 10^{10} L_{\odot,B}$ 
(e.g., Bahcall \& Soniera \cite{bs}).
In addition, placing the Milky Way
on the B-band Tully Fisher relation
of, e.g., Yasuda, Fukugita, \& Okamura \pcite{yfo}
with a Galactic circular velocity
$v_{\rm circ} = 220 \, {\rm km} {\rm s}^{-1}$
gives 
$L_{{\rm MW},B} = (1.9 \pm 0.6) \times 10^{10} L_{\odot,B}$.
It is encouraging that these estimates are consistent.
We will adopt a Milky Way B-band luminosity in the range 
\begin{equation}
L_{{\rm MW},B} = (1.3-2.5) \times 10^{10} \ L_{\odot B} \, .
\end{equation}

Thus, we can derive a Macho 
mass/Milky Way luminosity ratio of the 
Macho component of the Milky Way as 
\begin{equation}
\label{ml}
\Upsilon_{\rm Macho} \equiv 
  M_{\rm Macho}/L_{\rm MW}=(5.2-25) \, \msun/ \lsun \, .
\end{equation}
More precisely, this ratio includes only the Macho mass
by the LMC observations, i.e., out to 50 kpc.  Any additional
Macho population beyond this distance will increase
$\Upsilon_{\rm Macho}$ and strengthen the arguments below;
however, to be conservative we will adopt the
lower bound of Eq.\ (\ref{ml}) as the appropriate value.

{}From the ESO Slice Project
Redshift survey (Zucca \etal \cite{zuc}), 
the luminosity 
density of the Universe in the $B$ band is 
\begin{equation}
\label{phi}
{\cal L}_B = 1.9\times 10^{8} h \ \lsun \ \mpc^{-3}
\end{equation}
where the Hubble parameter 
$h=H_0/(100 \, {\rm km} \, {\rm sec}^{-1} \, \Mpc^{-1})$.  
If we assume that the $M/L$ which we defined for the Milky 
Way is typical of the Universe as a whole, 
then the universal mass density of 
Machos is 
\begin{equation}
\label{rho}
\rho_{\rm Macho} 
   = \Upsilon_{\rm Macho} \, {\cal L}_B
   = (1-5) \times 10^{9} h \ \msun \, \mpc^{-3} \, .
\end{equation}
The corresponding fraction of the critical density
$\rho_c \equiv 
3H_0^2/8 \pi G = 2.71 \times 10^{11} \, h^2 \, M_\odot \ \Mpc^{-3}$ is
\begin{equation}
\label{omega}
\omegam \equiv \rho_{\rm Macho}/ \rho_c = (0.0036-0.017) \, h^{-1} \, .
\end{equation}

Other alternative techniques for estimating $\omegam$ are possible
as well.
Thus far, we have explicitly assumed that Machos trace
visible light.  If Machos trace dark matter, 
then the total
Macho density can be derived by comparing the total mass of Machos
within 50 kpc to the total mass of dark matter
within 50 kpc (we consider the additional possibility of
biased galaxy formation later).  The Macho group has estimated
this fraction to be as high as 0.6; thus, $\Omega_{\rm Macho}$
could be as high as $0.6 \, \Omega$, where $\Omega$ without a subscript
refers to the total mass density of the universe.
Clearly this approach is very dependent on
the model of the Halo.  
We note that this estimate is higher than that of
Eq.\ (\ref{omega}) for all reasonable values of $\Omega$,
and so we will adopt the more conservative value of Eq.\ \pref{omega}
as our lower limit.

After completion of our calculations, we learned of
simultaneous and somewhat similar
calculations by
Fukugita, Hogan, \& Peebles \pcite{fhp}, who 
note that the Macho mass of Eq.\ \pref{macho_mass} is comparable to estimates
of the Galactic disk mass: $M_{\rm Macho} \sim 2 M_{\rm disk}$.  
This association of Machos with disks is similar to our
consideration of the case where only spiral galaxies have Machos
in them.  
They suggest extrapolation of this trend to
estimate the minimum Macho fraction of the critical
density and find a number which is quite similar
to ours.
\footnote{A discrepancy
of a factor of three between their results and ours stems
from a different disk luminosity implied by their paper.
Fukugita, Hogan, \& Peebles \pcite{fhp}
also consider Macho 
halo fractions approaching 100\%, and arrive at 
cosmic Macho densities which are quite large.}

In any case, we will now proceed to compare our $\omegam$
derived in Eq.\ \pref{omega} with the baryonic density in the universe,
$\omegab$, as determined by primordial nucleosynthesis.
The compact nature of the Machos strongly suggests they are
baryonic; we will assume this throughout this paper.  
Recently, the status of Big Bang
nucleosynthesis has been the subject of
intense discussion, prompted both by observations of deuterium
in high-redshift quasar absorption systems, and also by
a more careful examination of consistency and uncertainties
in the theory.
At present different groups report different D/H values,
e.g ``high'' D/H $\simeq 2 \times 10^{-4}$ 
(Carswell et al. \cite{cars94}; Songaila et al. \cite{song};
Webb et al. \cite{webb}; 
Carswell et al. \cite{cars96};
Wampler et al. \cite{wamp}); 
and ``low,'' D/H $= (3.4 \pm 0.3) \times 10^{-5}$
(Burles \& Tytler \cite{bt98a}; Burles \& Tytler \cite{bt98b}).
Using the ``low'' D/H values from Burles \& Tytler,
one obtains $\omegab= (0.019 \pm 0.0014) \ h^{-2}$ 
($1\sigma$ stat error).  To be conservative, we will
allow for a systematic error equal to the statistical uncertainty;
We will use the largest value in this range as our $1\sigma$ upper
limit: $\omegab \le 0.022$.
Our lower limit on $\omegab$ comes from consideration of observations
of other elements.  As emphasized by e.g., 
Fields et al.\ \pcite{fkot},
the most straightforward 
extrapolation of
primordial \he4\ and \li7\ from observations
is consistent with high but not low D/H and
gives $\omegab= 0.006^{+0.009}_{-0.001} \ h^{-2}$.

To conservatively allow for the full range of possibilities,
we will therefore adopt
\begin{equation}
\label{omegab}
\omegab= (0.005-0.022) \ h^{-2} \, .
\end{equation}

Note the different dependence on $h$ in equations (\ref{omega}) and 
(\ref{omegab}).  Despite this fact, we can still affirm that 
$\omegam$ and $\omegab$ are roughly comparable within this na\"{\i}ve 
calculation.  Thus, if the Galactic halo Macho
interpretation of the microlensing
results is correct,
Machos make up an important fraction of the baryonic matter 
of the Universe.  
Specifically, the central values in
eqs.\ (\ref{omega}) and (\ref{omegab}) give
$\omegam/\omegab \sim 0.7$.
However, the lower limit on this fraction is
considerably smaller and hence less restrictive.
Taking the lowest possible
value for $\omegam$ and the highest possible value for $\omegab$,
we see that
\begin{equation}
\label{comp}
{\omegam \over \omegab} \geq {1 \over 6} h
\end{equation}
so that
\begin{equation}
\label{comp2}
{\omegam \over \omegab} \geq \frac{1}{12} \, .
\end{equation}
For completeness, the corresponding upper limit is 
$\omegam/\omegab \leq 3 h$.  Note that if this
were the case (i.e., high $\omegam$, low $\omegab$),
then one would conclude that Machos are non-baryonic.

We believe our lower limit on the contribution of
Machos to the mass density of the universe to be quite
robust.
The only way that $\omegam$ could be lower than the numbers
presented in Eq.\ (\ref{omega}) would be if
the luminosity density observed by
Zucca et al. \pcite{zuc}\ 
were dominated by galaxies which do not contain Machos.
In other words, our Milky Way would have to be unrepresentative
of other galaxies in that it contains exceptionally many Machos.
However, the luminosity function observed by 
Zucca et al. \pcite{zuc}
has most of its light coming from galaxies similar to our own.
Hence the lower limit to $\omegam$ in Eq.\ (\ref{omega}) 
can be reduced by at most a factor of about five, as we will now show.
To see this we will 
adopt the
Zucca et al. \pcite{zuc}
luminosity function, which is well-fit by a Schechter function:
\begin{equation}
\label{schechter}
\phi(L) \, dL = \phi_* \, (L/L_*)^\alpha \exp(-L/L_*) \, dL/L_* \, ,
\end{equation}
where $\alpha = - 1.22^{+0.06}_{-0.07}$, and
$L_* = (1.1 \pm 0.07) \times 10^{10} \, h^{-2} \, L_{\odot,B}$.
Then, to obtain the fraction of Galaxies that do contain
Machos, we must integrate this luminosity function
over the range of luminosities that
we believe are relevant to galaxies that do contain
Machos.  The fraction of galaxies that contain Machos would then be
\begin{equation}
\label{lum_frac}
f_{\rm gal} = \frac { {\cal L}_{\rm Macho} } { \cal L }  = 
  \frac { \int_{L_1}^{L_2} \ dL \ L \ \phi(L) } 
        { \int_0^{\infty} \ dL \ L \ \phi(L) } \ \ .
\end{equation}

The choice of lower and upper limits of integration
($L_1$ and $L_2$ respectively)
is a matter of taste.
In the extreme case where only the Milky Way contains
Machos, the range of integration would be so narrow as to
obtain a negligible Macho fraction in the universe. 
However, a very narrow range is needed to
make this fraction small:  even if Machos only exist in 
Galaxies within one magnitude of the Milky Way
(i.e., $M_{{\rm MW},B} \pm 1 \, {\rm mag}$),
this still amounts to 26\% of the luminosity density
for $h=1/2$ or 50\% for $h=1$.
Of these galaxies roughly a fraction 0.65 are spiral
(Postman and Geller 1985).  Thus $\omegam$ in Eq.\ \pref{omega} should
be reduced by at most a factor of 0.17.

Regardless of the issue of which luminous galaxies contain Machos,
there remains a strong possibility that galaxy halos
suffer from cosmic biasing.  Namely,
it not all at clear that the mass-to-light ratio
determined at the scale of galaxy halos is representative of
the universe as a whole.
Indeed, density determinations at the scale of clusters
probably demand that mass does not trace light
at the halo scale.
If so, (and if $\Omega_0 \approx 1$)
then there could exist a population of
stillborn, dark galaxies or intracluster gas with baryons in them. 
One possibility is that
these dark galaxies contain no stars or Machos.
Then, in this case, the amount of baryonic
matter available to make Machos only becomes smaller;
i.e., $\omegam$ becomes an even larger fraction of
$\omegab - \Omega_{\rm dark}$, where $\Omega_{\rm dark}$ refers
to the dark galaxy component.  Alternatively, a second possibility
is that these dark galaxies
do contain Machos (e.g. left over from an early generation of star
formation).  In this case Machos trace the dark matter
and clearly the $\omegam$ in Eq.\ (\ref{omega}), obtained on 
the scale of galactic halos, is again
an underestimate of the true value.
In summary, we feel that the results of Eq.\ (\ref{omega}-\ref{comp2})
are hard to avoid.

\section{Comparison with the Lyman-${\bf \alpha}$ Forest}

We can compare the Macho contribution to other components
of the baryonic matter of the universe.  In particular,
measurements of the Lyman-$\alpha$ (\lya) forest absorption from 
intervening gas in the lines of sight to high-redshift
QSOs indicate that many, if not most, of the baryons
of the universe were in this forest at redshifts $z>$2.
It is hard to reconcile the large baryonic abundance
estimated for the \lya\ forest with $\omegam$
obtained previously (Gates, Gyuk, Holder, \& Turner \cite{gght}).  
Although measurements of the \lya\
forest only obtain the neutral column density, careful
estimates of the ionizing radiation can be made to 
obtain rough values for the total baryonic matter,
i.e. the sum of the neutral and ionized components,
in the \lya\ forest.  For the sum of these
two components,
Weinberg et al. \pcite{wmhk} estimate
\begin{equation}
\label{lyman}
\Omega_{\rm Ly\alpha} \sim 0.02 h^{-3/2} \, .
\end{equation}
This number is at present uncertain.  For example, it
assumes an understanding of the UV background responsible
for ionizing the IGM, and accurate determination
of the quasar flux decrement due to the neutral
hydrogen absorbers.  Despite these uncertainties,
we will use Eq.\ ({\ref{lyman}}) below
and examine the implications of this estimate.

We can now require that the sum of the Macho energy density
plus the \lya\ baryonic energy density do not
add up to a value in excess of the baryonic density from
nucleosynthesis:
\begin{equation}
\label{insist}
\omegam(z) + \omegalya (z) \leq \omegab \, ;
\end{equation}
this expression holds for any epoch $z$.  
Unfortunately, the observations of Machos and 
\lya\ systems are available for different epochs.
Thus, to compare the two one must assume
that there has not been a tradeoff of gas into
Machos between the era of the Lyman systems
($z \sim 2-3$) and the observation of the Machos at $z=0$.
That is, we assume that the Machos were formed
before the \lya\ systems. 

Although Eq.\ (\ref{insist}) offers a potentially
strong constraint, in practice the uncertainties in both 
$\omegalya$ and in $\omegab$ make a quantitative
comparison difficult.
Nevertheless, we will tentatively use
the numbers indicated above.
We then have 
\begin{equation}
\label{onehalf}
(\omegam = 0.007-0.04) + (\omegalya = 0.06) \leq (\omegab = 0.02 - 0.09)
\ \ \ {\rm for} \, h=1/2 \, ,
\end{equation}
and 
\begin{equation}
\label{one}
(\omegam = 0.004 - 0.02) + (\omegalya = 0.02) \leq (\omegab = 0.005 - 0.02)
\ \ \ {\rm for} \, h=1 \, .
\end{equation}
These equations can be satisfied, but
only if one uses the most favorable extremes
in both $\omegam$ and $\omegab$, i.e., for
the lowest possible values for $\omegam$
and the highest possible values for $\omegab$.

Recent measurements of Kirkman and Tytler \pcite{kt}
of the ionized component of a Lyman limit system
at z=3.3816 towards QSO HS 1422+2309
estimate an even larger value for the mass density in 
hot and highly ionized gas in the intergalactic
medium: $\Omega_{\rm hot} \sim 10^{-2} h^{-1}$.
If this estimate is correct, then
Eq.\ (\ref{insist}) becomes even more
difficult to satisfy.

One way to avoid this mass budget problem would
be to argue that the \lya\ baryons later
became Machos.  Then it would be inappropriate to
add the \lya\ plus Macho contributions in
comparing with $\omegab$, since the Machos would
be just part of the \lya\ baryons.  However,
the only way to do this would be to make the Machos
at a redshift after the \lya\ measurements were
made.  Since these measurements extend down to about
$z\sim 2-3$, the Machos would have to be made
at $z<2$.  However, this would be difficult to maneuver.
Stellar remnants could not have formed after redshift 2; we would
see the light from the stars in galaxy counts (Charlot and Silk \cite{cs})
and in the Hubble Deep Field (Loeb \cite{loeb}).

Until now we have only considered the contribution to the baryonic
abundance from the Machos themselves.  Below we will consider
the baryonic abundance of the progenitor stars as well, in the
case where the Machos are stellar remnants.  
When the progenitor baryons are added to the left hand side
of Eq.\ (\ref{insist}), this equation becomes harder to satisfy.
However, we wish to reiterate that measurements of $\Omega_{{\rm Ly} \alpha}$
are at present uncertain, so that it is possibly premature to
imply that Machos are at odds with the amount of baryons in 
the Ly$\alpha$ forest.

\section{On Stellar and Brown Dwarf Macho Candidates}
\label{sect:brown}

The arguments we presented in the previous sections
have been quite general, independent of the nature
of the Machos. In particular, they apply to 
stellar and brown dwarf Macho candidates.
Here we outline the existing stellar and substellar candidates. 
In the next section we discuss stellar remnants,
where the mass budget constraints become even more severe.

Before microlensing experiments reported their first results, the 
community had arrived at a general consensus that the
most plausible candidates for baryonic dark matter
in the halo of our Galaxy
were brown dwarfs (\eg Hegyi \& Olive \cite{ho})
or very low mass faint stars (red dwarfs).  The low mass tail of 
the mass function of halo stars was ill-known and it seemed natural that 
it would rise steeply enough to include large numbers of brown dwarfs (\eg 
Carr \cite{carr}).  
This conclusion was strengthened by the analysis of the first 
microlensing events towards the LMC 
(Alcock \etal \cite{macho:1yr}) which had a short 
time-scale (typically $\leq 40$ days) suggesting that a decent fraction of the halo was composed of 
objects with mass about $0.1 \msun$, very plausibly brown dwarfs.

However, a significant population of red dwarfs (faint low mass 
hydrogen burning stars) has been ruled out by Hubble Space
Telescope data and by parallax data, which simply did
not see large numbers of these objects 
(Bahcall, Flynn, Gould \&  Kirhakos \cite{bfgk};
Graff \& Freese \cite{gf96a,gf96b}; 
Dahn et al. \cite{dahn})  In addition,
measurements of the mass function of halo 
stars showed that the mass function does not rise fast enough (towards low 
mass) to allow a significant population of brown dwarfs 
(M\'{e}ra, Chabrier \& Schaeffer \cite{mcs}; 
Graff \& Freese \cite{gf96b}).
In fact, we found that red dwarfs and brown dwarfs do not
constitute more than $\sim 3$\% of the mass density of the
Halo of the Galaxy.  These bounds follow from the simple
assumption that halo red dwarfs are distributed isotropically.
Kerins (\cite{eammon1,eammon2}) discussed how these bounds on the mass density
are weakened if halo red dwarfs are clustered.
Kerins has also suggested (reported at the 1998 International
Workshop on Gravitational Microlensing) that brown dwarf candidates might
remain if their orbits keep them
predominantly at large radii (far from the center of the Galaxy).

In addition, 
all subsequent microlensing events had time-scales longer than the first LMC 
microlensing event.  After two years, the time scales of the events were 
long enough that the estimated mass of Machos grew to $0.5 \msun$, clearly 
inconsistent with a brown dwarf population (which must have masses 
$\la 0.1 \msun$; 
Alcock \etal \cite{macho:2yr}).  Now, after four years of Macho 
group runs, events continue to have the large time scales consistent with 
white dwarfs (D. Bennett, reported at the 1997 Notre Dame 
microlensing 
conference).  

There are still several other possible stellar or substellar
explanations for the microlensing 
results:  A brown dwarf halo in which the component brown dwarfs are on 
radial orbits would mimic the long time scale events measured by the 
microlensing groups (Evans \cite{evans}).  
A population of ordinary stars lying on 
the line of sight to the LMC could also explain the microlensing signal.  
Such a population could be due to a tidal tail of the LMC 
(Zhao \cite{zhao}; 
Zaritsky \& Lin \cite{zl}; but see Gould \cite{gould}; 
Alcock \etal \cite{macho:nodwf}; Beaulieu \& 
Sackett \cite{bsac}).  However, if we examine only microlensing evidence and 
Galactic dynamics, white dwarfs are the most promising 
candidate Machos.

\section{Machos as Stellar Remnants: the Mass Budget
Constraints from the Macho Progenitors}

\noindent
In this section we will investigate the possibility
that Machos are stellar remnants: white dwarfs, neutron
stars, or black holes.  
We will now define and
relate a few useful quantities:  $m$ will refer to the
initial mass of the main sequence star,
which is the progenitor of a remnant having mass
$\mrem(m)$.   Unless otherwise specified, all masses
are in solar units.

For the Macho mass density contribution discussed above,
we here estimate the contribution of Macho progenitors to
the mass density of the Universe.  
We find it useful to 
define $r$ to be the ratio of the total mass in Machos to the
total mass in progenitors. 
To do this requires specification of an initial mass function
for the Macho progenitors; we will denote this IMF as
$\imfm(m) = dN_{\star}/dm$.
The IMF we choose will be different depending
on whether the dominant remnant component is 
a white dwarf or a neutron star.
Given a Macho progenitor IMF, we can define
\begin{equation}
\label{eq:rbar}
\rbar  =   \frac {\int_{1 \msun}^\infty dm \ \mrem(m) \, \imfm(m)} 
             {\int_{0}^\infty dm \ m \, \imfm(m) } \ .
\end{equation}
Equation \pref{eq:rbar} is closely related
to the usual gas return fraction $R$ by $\rbar = 1-R$.
Here $R$ is the fraction of the progenitor mass returned as gas
(i.e., not swept up into the remnant).
Our lower limit in the integral in the numerator
arises because stars with mass $M \la 1\msun$ 
will not have completed their stellar evolution in a Hubble time.

\subsection{The Case of White Dwarfs}

In this section, we focus on the case where
the Machos are mostly white dwarfs with a smattering
of neutron stars.  This choice allows us to be concrete
and to evaluate the integrals above.  As mentioned previously,
white dwarfs appear to be the best fit candidates to the
mass estimates of the Macho group.  We will show that the 
ratio of the mass in Machos to progenitors can be at most 
$r<1/2$ (e.g., $r=1/4$ for the log-normal mass function described below),
so that (in a single burst model)
the progenitor mass density is at least twice as large
as the Macho mass density we have already computed.

{\it Initial Mass Function:} 
The nature and plausibility of the white dwarf progenitor 
mass function (for brevity, ``white dwarf IMF'')
is a central issue for 
(and a major source of objections to)  a white dwarf halo scenario.
Unfortunately, it is always difficult to derive the 
IMF empirically, as emphasized 
in the excellent review by Scalo \pcite{scalo}.
An IMF of the usual Salpeter \cite{salp})
type, $\imfm(m) \propto m^{-2.35}$ extending to 
$m_{\rm min} \sim 0.1 \msun$,
is not appropriate, as it would imply a gross overabundance of low mass
stars still in the halo.  Any white dwarf IMF must 
be sharply different from any observationally inferred IMF.
This difference 
justifiably strikes many as being a sign of fine tuning required for a white
dwarf halo model.  However, one should bear in mind
the significant 
model-dependences inherent in deriving any IMF. Also,
the star formation theory 
of Adams and Fatuzzo \pcite{af} predicts that a zero metallicity
primordial gas, such as was present at the time of star
formation in the Halo, 
would form higher mass stars than a nonzero
metallicity gas, which formed all the familiar stars.

The dearth of stars in the dark Halo provides a lower limit on the
mass of the white dwarf progenitors.  
The most solid constraint simply demands that 
the white dwarf progenitors all lived less
than the age of the universe
($t_0 \simeq 11-17$ Gyr for most plausible cosmologies).
With the mass-lifetime
relation of, e.g., Scalo \pcite{scalo}, this translates
to a hard lower cutoff mass of $m_{\rm low} \simeq 0.85-0.95 \msun$.
In addition, the white dwarfs themselves have to cool off
in the age of the universe in order to explain why they
are not seen in a recent search of the Hubble Deep Field
(Flynn, Bahcall, and Gould 1996) and in a ground based
search by Liebert, Dahn, and Monet (1988).
Looking at Figure 6 in Graff, Laughlin, and Freese (1997),
one can see that white dwarfs with progenitors lighter
than 1$M_\odot$ would not have been able to cool off enough in 18 Gyr
to avoid detection in these two surveys,
unless white dwarfs constitute less than 4$\times 10^{-4}
M_\odot$ pc$^{-3}$, roughly 4\% of the Halo.
The precise lower limit on the progenitor mass in order
for the white dwarfs to have cooled sufficiently depends
on the correct number for the age of the universe as well
as on the Halo model.  For example,
white dwarfs that constitute at least $10^{-3} \msun \, {\rm pc}^{-3}$
would have cooled off sufficiently in 13 Gyr if their
progenitors were more massive than 1.5 $\msun$.
Because of these uncertainties, we will consider a range
of lower limits on the progenitor mass in our calculations below.

The upper limit to the white dwarf IMF is not as strong.
By assumption, we wish white dwarfs to be the most numerous
halo remnants.  Thus we want an  IMF with most of the
progenitors  less massive than
$m_{\rm WD,max} \sim 8 \msun$,
since progenitor stars
heavier than $\sim 8 M_\odot$ explode as supernovae
and become neutron stars.
Adams and Laughlin \pcite{al} take this mass as an
upper bound, since they view neutron stars as unwanted.
However, having {\em some} high mass stars
is permissible and even desirable (Miralda-Escud\'e \& Rees \cite{mr};
Gnedin \& Ostriker \cite{go}):
high redshift supernovae
are required to produce the metals $Z \sim 10^{-2} Z_\odot$
seen in quasar absorption
lines of quasars at redshift $z \la 5$.  The supernovae may
also provide a Galactic wind to eliminate excess metals and mass
(\S \ref{sect:wind}).  
The upper limit is thus model-dependent, 
aside from the basic requirement of white dwarf (rather
than neutron star) dominance.

Within these general mass limits, there remains considerable
freedom for the form of the white dwarf IMF.
Fortunately, we find that
one can already make interesting statements 
about the white dwarf progenitors just using these
limits.  Further constraints must rely on
theoretically derived IMFs, or on the details of the
halo white dwarf scenario one adopts.
For example,  
Adams and Laughlin \pcite{al}  use a log-normal
mass function motivated by Adams \& Fatuzzo's \pcite{af}
theory of the IMF:
\begin{equation}
\label{lognormal}
\ln {\imfm}(\ln m) = A - {1 \over 2 \sigbar^2}
\Bigl\{ \ln \bigl[ m / m_C \bigr] \Bigr\}^2 \, . 
\end{equation}
The parameter $A$ sets the overall normalization. The mass scale
$m_C$ (which determines the center of the distribution) and the
effective width $\sigbar$ of the distribution are set by the
star-forming conditions which gave rise to the present day population of
remnants. For illustration, we adopt their
white dwarf IMF parameters $m_C=2.3 M_{\odot}$
and $\sigbar=0.44$, which imply warm, uniform star-forming conditions
for the progenitor population.  These parameters saturate the twin constraints
required by the low-mass and high-mass tails of the IMF, as discussed
by Adams \& Laughlin \pcite{al}, i.e., this IMF is as wide as possible
if supernovae are to be excluded (but see above).

{\it Initial/Final Mass Relation:}
The relation between the mass of a progenitor star and 
the mass of its resultant white dwarf relies on an
(imperfect) understanding of mass loss from red giants.  We use
the results of 
Van den Hoek \& Groenewegen \pcite{vdhg}; these
are consistent with the results of Iben \& Tutukov \pcite{it}.
At the progenitor mass limits of interest, we have white dwarf masses
$m_{\rm WD}(1 \msun) = 0.55 \msun$, and
$m_{\rm WD}(8 \msun) = 1.2 \msun$.
The key point we will use below is that the ``remnant
fraction'' relative to the progenitor mass, $\mrem/m$, 
is a monotonically decreasing function of $m$.

{\it White Dwarf Progenitor Mass Budget:}
Our goal is to estimate the contribution of the white
dwarf progenitors to the mass density of the universe.
Given $\omegam$ from the previous sections, we now need
to compute the remnant fraction \rbar\ of Eq.\
\pref{eq:rbar}.  Despite the freedom in choosing the white dwarf IMF
as discussed above, interesting and general limits to \rbar\ still emerge.
We will bound the possible values for \rbar\ in the following way.
Consider the remnant fraction at a given mass,
$r(m) = \mrem(m)/m$.  One can rewrite
Eq.\ \pref{eq:rbar} as 
\begin{equation}
\label{rewrite}
\rbar = \frac{\int dm \ r(m) \, m \, \imfm(m)}
             {\int dm \ m \, \imfm(m)} \, .
\end{equation}
In this form one can see than $\rbar$ is a weighted average of $r(m)$
where $m \, \imfm(m)$ is the weighting function.  Then if $r(m)$ has
a maximum (minimum) at some $m$, then the weighted average $\rbar$ is always
greater than (less than) this extremum.
But $\mrem(m)/m$ is maximized for the lowest allowed progenitor mass,
$m_{\rm min}$.  If $m_{\rm min} = (1,1.5,2) \msun$, then
correspondingly $r_{\rm max} = (0.55,0.38,0.33)$, i.e.,
white dwarfs are at most about half of the progenitor mass.
Similar arguments apply to finding a minimum value for \rbar;
we find $\rbar > r_{\rm min} = 0.15$, for
$m_{\rm max} = 8 \msun$.  The most conservative case is the
one that gives the lowest progenitor mass density $\Omega_*$,
as would be obtained with $r_{\rm max}$,
the largest possible value for \rbar.
We have found the
range of possible values for \rbar, and we see that
$M_\star = M_{\rm Macho}/\rbar \geq 2  M_{\rm Macho}$. 
Thus, for every unit mass of Machos formed, 
at least as much mass is 
ejected as gas as swept into remnant Machos.  For comparison,
if we use the Adams \& Laughlin \pcite{al} 
log-normal IMF in Eq.\ (\ref{lognormal})
we find $R = 0.75$ and $\rbar = 0.25$, which implies that three
times as much mass is ejected as gas as is swept into 
remnants.  

We note that Fields, Mathews \& Schramm \pcite{fms} used an
IMF of the same log-normal form as Eq.\ \pref{lognormal}, but
with a larger width:  $\sigbar = 1.6$.  This gave a return 
fraction of $R=0.375$.  In their model, the quantity of gas ejected was 
only about half the total Macho mass.  Thus, for a given
total mass in Machos, the Fields, Mathews and Schramm model
produced only one fifth as much returned gas as would be produced 
by the canonical Adams \& Laughlin mass function.
This comes about largely due to the assumption that
the white dwarf IMF extends above $8 \msun$ to include
a significant fraction of massive stars that are above the
Bethe \& Brown \pcite{bb} $18 \msun$ cutoff for direct
black hole formation.  While these black holes
are rare in number compared to white dwarfs,
in fact they carry a significant fraction of the mass
from the Fields, Mathews \& Schramm \pcite{fms} IMF.

Having derived constraints on $\rbar$, we must still specify 
more information about the white dwarf formation to
constrain the progenitor mass.  Specifically, we must
address the issue of reprocessing of the gas ejected
when the white dwarfs are formed.

{\it Single Burst Model:}
If we assume that Machos were born in a single burst, then that portion of 
the mass of the progenitor stars which does not go into the remnants 
cannot 
be used for subsequent formation of Machos.  
Then we can immediately infer 
the progenitor density $\rho_\star$
from the Macho density via Eq.\ \pref{eq:rbar}:
$\rho_\star = \rbar^{-1} \rho_{\rm Macho}$.  
Thus, the upper limit $r_{\rm max} = 0.55$ 
(for $m_{\rm min} = 1 \msun$)
now gives a lower limit
to the progenitor density,
\begin{equation}
\label{rhostar}
\rho_\star \ge \frac{1}{r_{\rm max}} \rho_{\rm Macho} 
   = (2 - 9) \times 10^{9} h \ \msun \  \mpc^{-3}
\end{equation}
which corresponds to
\begin{equation}
\label{Omegastar}
\Omega_\star \ge (0.007-0.034) h^{-1} \, .
\end{equation}
If we adopt $m_{\rm min} = 1.5 \msun$ as suggested by
white dwarf cooling arguments, this bound increases
by 30\%.  
If the white dwarf IMF is specified, these expressions
become equalities.  
For example, with the IMF of Eq.\ (\ref{lognormal}), 
we have $R = 0.75$ and $\rbar = 0.25$ so that
the total mass density 
of the progenitors was roughly four times higher than the mass
density of the remnant Machos,
and $\Omega_\star = (0.016 - 0.08) h^{-1}$.

The highest $\omegab$ from
BBN in Eq.\ (\ref{omegab}) would 
be consistent with the lowest $\Omega_\star$ from the simple
extrapolation above if
$h<0.9$.  Thus there is a range of overlap between the two estimates,
if one takes the low estimates of D/H and low total
Macho mass $\omegam$.
Since these ranges for the Hubble constant and D/H are, if anything,
currently favored, this situation is not so unattractive.  
Of course, the mass budget must include all
types of baryons, in particular, those in the IGM (Eq.\ \ref{insist}).
However, the relationship between $\Omega_\star$ and $\Omega_{\rm IGM}$
depends on the scenario for white dwarf formation;
if the white dwarf progenitors are born at high redshift,
the the net ejecta $\Omega_\star - \omegam$ becomes part of
the IGM, and thus should not be double counted in the
mass budget (but see also nucleosynthesis issues 
in \S \ref{sect:carbon}).

{\it Recycling}:  As an alternative to the model discussed
above in which all of the Machos were created in a single
burst, there is the possibility of recycling.
In the recycling scenario, much of the mass let out by high mass Macho 
progenitors 
is incorporated in the next generation of stars.  This would allow us to 
make more Machos with less gas, and have less intergalactic gas left over 
too.  However, packing most of the gas into white dwarfs through several 
star formation generations could create too much helium (Ryu, Olive \& Silk 
\cite{ros}).  In addition, as discussed earlier,
$\Omega_{\rm IGM} \sim 0.02 h^{-3/2}$ of gas should 
be left over after the Macho formation epoch to form the intergalactic 
medium (Weinberg, Miralda-Escud\'e, Hernquist \& Katz \cite{wmhk}).
The chemical evolution of a recycling scenario should be
considered further.
The fact that a large
fraction of baryons must be cycled through the Machos and
their progenitors must be taken into account when attempting
to dilute the chemical by-products of such a large population of
intermediate mass stars.  Further discussion of this
issue follows shortly.

\subsection{Galactic Winds}
\label{sect:wind}

The white dwarf progenitor stars
return most of their mass
in their ejecta, i.e., planetary nebulae 
composed of processed material.
Both the mass and the composition of the material
are potential problems.  As we have emphasized, 
the cosmic Macho mass budget is a serious issue; 
here we see that it is significant
even when one considers only the Milky Way.

One can use $\rbar$ to derive not only cosmological limits, but also
local ones.  Given the $M_{\rm Macho}$ of Eq.\ \pref{macho_mass}, 
a burst model requires the total mass of progenitors in the Galactic Halo
(out to 50 kpc) to have been at least twice the total mass in remnant
white dwarfs, i.e.,
$M_\star \ge M_{\rm Macho}/r_{\rm max} = (2.4 - 5.8) \times 10^{11} \msun$.
The gas that is ejected by the Macho
progenitors is collisional and tends to fall into the Disk
of the Galaxy.
But the mass of the ejected gas $M_{\rm gas}
= M_\star - M_{\rm Macho} \sim M_{\rm Macho}$
is at least as large as the mass ($\sim 10^{11} M_\odot$)
of the Disk and Spheroid of the
Milky Way combined.  
For a value of \rbar\ less than $r_{\rm max}$, the gas
ejected by the Macho progenitors exceeds the mass of the Disk
and Spheroid. 
Thus the Galaxy's baryonic mass budget---including Machos---immediately 
demands that 
some of the ejecta be {\it removed} from the Galaxy.

This requirement for outflow is intensified when 
one considers the composition of the stellar ejecta.
It will be void of deuterium, and will include large amounts
of the nucleosynthesis products of $(1-8) \msun$ white
dwarf progenitors, notably:  helium, carbon, and nitrogen 
(and possibly s-process material).
To reduce these abundances to acceptably low
levels requires
some degree of {\it dilution}:  mixing the ejecta
with material that has not gone into stars.
In other words, the efficiency of star formation
(in terms of mass into stars compared to total available
mass) must be small.  This implies that there
is additional gas mass associated with the white dwarfs,
but not coming from the white dwarfs or their progenitors.
This mass too must
be partially (mostly!) removed to respect the
observed disk and bulge baryon budget.
Thus we see that  a ``closed box''
model of white dwarf Macho formation fails for the Galaxy.  Some means of
gas outflow is needed.

A possible means of removing these excess baryons
is a Galactic wind.  Indeed, as pointed out
by Fields, Mathews, \& Schramm \pcite{fms}, 
such a wind may be a virtue, as hot gas containing
metals is ubiquitous in the universe, seen in
galaxy clusters and groups, and present as an ionized
intergalactic medium that dominates the observed
neutral \lya\ forest.  Thus, it seems mandatory
that many galaxies do manage to shed hot, processed material.  
Of course,
it is considerably more difficult to construct a 
detailed scenario
which quantitatively shows that
the gas temperatures and compositions, generated concurrently
with white dwarf Machos, are consistent with the observations.  
There are serious problems
in matching the carbon abundances arising from white
dwarf progenitors with the carbon abundances in the 
\lya\ forest and elsewhere, as we discuss in \S \ref{sect:carbon}.

Before constructing a detailed model, 
one can deduce several 
necessary features of a
white dwarf + wind scenario.
(1)  Outflow presumably requires massive stars as
an energy source; these must be included in the
halo IMF.  They will thus contribute supernova products
to the wind composition.  Indeed, 
such elements are present in observed,
wind-ejected material:  intracluster X-ray gas
contains iron, oxygen, and other
$\alpha$-chain elements
coming from supernovae (e.g., Mushotzky et al.\ \cite{mush}).  
(2) For the wind to be effective, it must remove
material ejected from a large fraction of the planetary 
nebulae formed.  But the winds are driven by shorter-lived,
massive stars which explode as supernovae.  
This implies that the progenitors cannot have
all formed simultaneously, so that some massive stars are left 
to explode as the planetaries appear.  Such an arrangement
requires an extended burst of star formation,
lasting roughly $t_{\rm burst} \sim \tau(m_{\rm min})$,
the age of the longest lived white dwarf progenitor.  This could
be as long as 1 Gyr (for $m_{\rm min} = 2 \msun$),
unless $m_{\rm min}$ is larger (reducing the timescale but raising
the total progenitor mass that must then be removed).
(3)  It is clear solely from
considerations of the implied progenitor mass budget, and
from the further need to dilute the ejecta composition, 
that white dwarf halos cannot be where most
of the universe's baryons reside.  Thus, white dwarf halos,
if they exist, would be only the tip of the 
baryonic iceberg---i.e., the white dwarfs themselves would
represent only a minority of the baryons needed
to accommodate their formation and associated pollution.

Of course, we have only considered the need for
outflow in the burst approximation, in which
all progenitors are formed before any have died
and returned their ejecta.  In cases with
recycling, the total mass budget is reduced
since the ejecta of successive generations become
grist for new stars.  However, where recycling
alleviates the mass problem, it exacerbates
chemical composition problems.  That is, 
the contamination of the gas due to
stellar nucleosynthesis processing is compounded
as the gas is recycled.  Thus abundance constraints
(\S \ref{sect:carbon})
work to require dilution and to limit recycling,
at odds with the need for baryonic parsimony.
The result is that outflow is still needed even
in a recycling model.

\subsection{On Carbon}
\label{sect:carbon}
The issue of carbon (Gibson \& Mould \cite{gm})
produced by white dwarf progenitors is complex.
We present here a simple estimate of the difficulties of
reconciling the carbon production with the baryonic abundance
of Machos.  Stellar carbon yields for zero 
metallicity stars are quite uncertain.
Still, according to the Van den Hoek \& Groenewegen (1997) yields, a star of mass 
2.5 will produce $1.26 \msun$ of ejecta of which $0.012 \msun$ is carbon, 
for an ejected mass fraction of $10^{-2}$.  In comparison, the sun has a 
carbon enrichment of $4.4\times 10^{-3}$.  In other words, the ejecta of a 
typical intermediate mass star has about twice the 
solar enrichment of carbon.  
If a substantial fraction of all 
baryons pass through intermediate mass stars, the carbon abundance in this 
model will be near solar.

It is possible (although not likely) that carbon never leaves
the white dwarf progenitors, 
so that carbon overproduction is not a problem
(Chabrier, private communication).
Carbon is produced exclusively in the stellar core.  
In order to be ejected, carbon must convect to the outer layers in 
the ``dredge up'' process.  Since convection is less efficient in a zero 
metallicity star, it is possible that no carbon would be ejected in a 
primordial star.  In that case, it would be impossible to place limits on 
the density of white dwarfs using carbon abundances. 
For the remainder of this section, we will assume
that carbon does leave the white dwarf progenitor stars.

Then overproduction of carbon can be a serious problem,
as emphasized by Gibson \& Mould \pcite{gm}.
They noted that stars in our galactic halo have carbon 
abundance in the range $10^{-4}-10^{-2}$ solar, 
and argued that the gas 
which formed these stars cannot have been polluted by the ejecta of a 
large population of white dwarfs.
The galactic winds discussed in the previous section could
remove carbon from the star forming regions and mix it throughout the 
universe.  

However, carbon abundances in intermediate redshift 
\lya\ forest lines have recently been measured to be 
quite low.
Carbon is indeed present, but only at the
$\sim 10^{-2}$ solar level,
(Songaila \& Cowie \cite{sc}) for \lya\ systems at $z \sim 3$
with column densities $N \ge 3 \times 10^{15} \, {\rm cm}^{-2}$.
Furthermore, in 
an ensemble average of systems 
within the redshift interval $2.2 \le z \le 3.6$,
with lower column densities 
($10^{13.5} \, {\rm cm}^{-2} \le N \le 10^{14} \, {\rm cm}^{-2}$),
the mean C/H drops to $\sim 10^{-3.5}$ solar
(Lu, Sargent, Barlow, \& Rauch \cite{lsbr}).
One can immediately infer that, however carbon is produced
at high redshift, the sources do not enrich all material
uniformly.  Any carbon that {\em had} been produced more
uniformly prior to these observations (i.e., at still
higher redshift) cannot
have been made above the $10^{-3.5}$ solar level.

The forest lines discussed in these references 
sample the high-$z$ neutral intergalactic medium.
If we assume that the nucleosynthesis products of
the white dwarf progenitors do not avoid the neutral medium,
then these observations 
offer strong constraints on scenarios for
ubiquitous white dwarf formation.
We will use the more conservative
$10^{-2}$ solar values, which provide the weaker constraints.
These values are quite plausible as well, since the higher
column density systems are more typical of regions
of high star formation, wherein white dwarf progenitors
might be formed.  
In order to maintain carbon abundances as low as $10^{-2}$ solar, only about 
$10^{-2}$ of all baryons can have passed through the intermediate mass 
stars that were the predecessors of Machos.  Such a fraction can barely
be accommodated by our results in section (\ref{naive})  
for the remnant density predicted from our extrapolation 
of the Macho group results, and would be in conflict with
$\Omega_\star$ in the case of a single burst of star formation.
The simple discussion we have presented here has considered only
the simplest (but most plausible) white dwarf formation and evolution
scenario.  More detailed investigation is required.  However,
we stress the difficulty of reconciling the Macho mass budget with
the accompanying carbon production in the case of white dwarfs.

\subsection{Neutron Stars}
The first issue raised by neutron star Macho candidates
is their compatibility with the microlensing results.
Neutron stars ($\sim 1.5 \msun$) and
stellar black holes ($\ga 1.5 \msun$)
are more massive objects, so that one would typically expect longer
lensing timescales than what is currently observed in
the microlensing experiments (best fit to $\sim 0.5 \msun$).
As discussed by Venkatesan, Olinto, \& Truran \pcite{vot},
one must posit 
(1) that the microlensing events observed
thus far only detect the low-mass tail of a
distribution that includes significantly more massive objects; and 
(2) that
as the experiments continue to take measurements, longer
timescale events should begin to be seen.
In this regard, it is intriguing that the
first SMC results (Palanque-Delabrouille et al.\ \cite{eros:smc};
Alcock et al.\ \cite{macho:smc}) suggest lensing masses
of order $\sim 2 \msun$.  
If indeed the present microlensing
results miss a higher mass lensing population, then
the halo Macho mass has been underestimated; thus
the baryonic budget constraints we have derived would  be
strengthened.

Even with the Galactic Macho mass we adopt in Eq.\ \pref{macho_mass},
the central issues for neutron stars
are the baryonic budget, and the chemical composition
of the stellar ejecta.   One expects
the progenitor mass requirements to be significantly
worse than for white dwarfs, since
for neutron stars, $r(m) = m_{\rm rem}/m \le  m_{\rm NS}/m_{\rm SN,min} 
\sim 1.5\msun/8\msun = 0.2$;
thus the mass density in progenitors must be at least 5 times
the neutron star mass density (cf.\ Eq.\ \ref{rhostar}),
and consistency with the upper bounds on $\omegab$ become hard
to satisfy.
However, Venkatesan, Olinto, \& Truran \pcite{vot} note
that this limit is reduced in some scenarios of stellar
black hole formation.  The models they consider
have remnant masses that smoothly increase with the progenitor mass;
to give a rough example of the effect, we note
that allowing ejecta-free black holes to form
at masses $> 30 M_\odot$, increases the mean remnant mass
so that $\rbar = 0.24-0.36$.
The composition of the ejecta will be highly metal rich,
in a situation similar to the carbon problem with white dwarfs.
Here, however, the higher yields are
also directly associated with
explosive heating that could drive winds.

Venkatesan, Olinto, \& Truran \pcite{vot}
discuss a scenario wherein a burst of early, massive 
($\ga 10 \msun$) star formation
leads to neutron star, black hole, and metal production.
The metallicity of the ejecta is typically much larger than solar,
with the best case being a zero initial metallicity model,
which has ejecta abundances
of 1.3 times solar.  
Venkatesan, Olinto, \& Truran \pcite{vot}
compute the mass dilution needed to reduce the 
mean metallicity to solar,
which is at the high end of the
metallicities found in
intracluster and intragroup gas.
The total initial mass needed
for this dilution is found to be compatible with $\omegab$ if
most baryons participated either in this star formation burst or
in the associated dilution.  Furthermore, most of the baryons end up
in intragroup hot gas, so the Local Group mass budget constraints could
be met as well.  However, assuming this evolution history is typical,
it is not clear how to reconcile this
scenario with the IGM abundances as determined by
the \lya\ absorption systems.  As discussed in \S \ref{sect:carbon},
these systems show typical abundances at least two orders of magnitude below
solar.  If one requires that the massive star ejecta be diluted to
this level, then the baryonic requirements become prohibitive.
More baryons are required to dilute the ejecta than are
allowed by nucleosynthesis.

\section{Conclusions}

If Machos are indeed found in halos of galaxies like our own,
we have found that the cosmological mass budget for Machos
requires $\omegam/ \omegab \geq {1 \over 6} h f_{\rm gal}$,
where $f_{\rm gal}$ is the fraction of galaxies that contain Machos,
and quite possibly $\omegam \approx \omegab$.  
Thus a stellar explanation of the
microlensing events requires that a significant
fraction of baryons cycled through Machos and their
progenitors. If the Machos are
white dwarfs that arose from a single burst of star
formation, we have found that the contribution of the progenitors
to the mass density of the universe is at least a factor of two higher,
probably more like three or four.  
We have made a comparison of $\omegab$ with
the combined baryonic component of $\omegam$  
and the baryons in the \lya\
forest, and  found that the values
can be compatible only for the extreme values of the parameters.
However, measurements of $\Omega_{{\rm Ly} \alpha}$
are at present uncertain, so that it is perhaps premature to
imply that Machos are at odds with the amount of baryons in 
the Ly$\alpha$ forest.
In addition, we have stressed the difficulty in reconciling the Macho mass
budget with the accompanying carbon production in the case of white
dwarfs.  The overproduction of carbon by the white dwarf progenitors
can be diluted in principle, but this dilution would require
even more baryons that have not gone into stars.  
At least in the simplest scenario, 
in order not to conflict with the upper bounds on $\omegab$,
this would require an $\omegam$ slightly smaller
than our lower limits from extrapolating the Macho results.
Only 10$^{-2}$ of all baryons can have passed through
the white dwarf progenitors, a fraction that is in conflict
with our results for $\Omega_\star$.

There is a certain amount of irony in the following:
There is as yet no detection of nonbaryonic dark matter,       
yet there is an abundance
of attractive candidate particles and galaxy formation
scenarios for these particles.
Here, on the other hand, we have been taking seriously
the possibility that a baryonic dark matter component
has been detected as Machos, yet no 
candidates seem able to convincingly explain both
local and cosmological observations,
though white dwarf candidates still survive at the limits
of the uncertainties.  Should stellar candidates be
definitively ruled out, this would be
a very interesting state of affairs-----indeed
more so than if the Halo is made of brown dwarfs or even
white dwarfs.  If it becomes clear that the stellar
or remnant contributions
to the mass density of the Halo are small, then
either (1) Machos are insignificant in the Halo, which means the Halo
is made of non-baryonic dark matter, or
(2) Machos are significant in the Halo, but
they are in an exotic form of baryons
or are not baryonic (e.g., primordial black holes).
Either way, the halo could well be made of very different
stuff than we are.  At present, however, 
a stellar explanation (including white dwarfs) 
for Machos survives as a possibility, albeit with the difficulties
pointed out in the paper.

\acknowledgements
We thank Elisabeth Vangioni-Flam, 
Grant Mathews, Scott Burles, Joe Silk, Julien Devriendt, Michel Cass\'e,
Jim Truran (who warned us from the beginning that carbon
overproduction would be a serious problem for white dwarfs), 
Sean Scully, and Dave Spergel for helpful
discussions.  We especially wish to thank
Dave Schramm, whom we already miss as a friend, colleague, and mentor,
and without whom the field of cosmology will not be the same.
We are grateful for the
hospitality of the Aspen Center for Physics,
where part of this work was done.
DG acknowledges the financial support
of the French Ministry of Foreign Affairs' Bourse Chateaubriand.
KF acknowledges support from the DOE at the
University of Michigan.  
The work of BDF was
supported in part by
DoE grant DE-FG02-94ER-40823.

\end{document}